\documentclass[10pt,letterpaper,onecolumn,accepted=2021-01-04]{quantumarticle}
\pdfoutput=1
\usepackage[utf8]{inputenc}
\usepackage[english]{babel}
\usepackage[T1]{fontenc}
\usepackage{mathtools, amssymb}
\usepackage[margin=1in]{geometry}
\usepackage{graphicx, xcolor, soul}

\usepackage{authblk}
\usepackage{makecell}
\usepackage{multirow}

\usepackage[numbers,sort&compress]{natbib}

\definecolor{mypurple}{RGB}{164,64,214}
\definecolor{mypurple2}{RGB}{170,0,255}
\definecolor{mycyan}{RGB}{0, 191, 255}
\definecolor{myred}{RGB}{255, 0, 85}
\definecolor{mypink}{RGB}{255, 0, 213}
\definecolor{qmagenta}{RGB}{199, 0, 57}
\definecolor{qblue}{RGB}{37, 81, 127}

\usepackage{tikz}
\usepackage[hyperindex,breaklinks]{hyperref}
\hypersetup{
    colorlinks=true,
    linkcolor=qmagenta,
    filecolor=qmagenta,      
}
\usepackage[all]{hypcap}

\newcommand{\ket}[1]{| #1 \rangle}
\newcommand{\bra}[1]{\langle #1 |}

\newcommand{\bsquare}{\scalebox{0.7}{$\blacksquare$}}

\newcommand{\ceil}[1]{\lceil #1 \rceil}
\newcommand{\floor}[1]{\lfloor #1 \rfloor}

\title{Exponentially faster implementations of Select($H$) for fermionic Hamiltonians}
\author{Kianna Wan}
\affiliation{Stanford Institute for Theoretical Physics, Stanford University, Stanford, CA 94305, USA}
\affiliation{PsiQuantum, Palo Alto, CA 94304, USA}
\email{kianna@stanford.edu}
\orcid{0000-0002-1147-6528}
\thanks{This research was completed during an internship at PsiQuantum.}

\usepackage{qcircuit}
\usepackage{pbox}

\newcommand{\injZ}{\pbox{3em}{\textsc{Inj}\\$(Z)$}}
\newcommand{\injselQ}{\pbox{3em}{\textsc{Inj-}\\\textsc{Sel}\\$(Q)$}}
\newcommand{\injselP}{\pbox{3em}{\textsc{Inj-}\\\textsc{Sel}\\$(P)$}}

\begin{document}


\maketitle

\begin{abstract}
We present a simple but general framework for constructing quantum circuits that implement the multiply-controlled unitary $\textsc{Select}(H) \coloneqq \sum_\ell \ket{\ell}\bra{\ell}\otimes H_\ell$, where $H = \sum_\ell H_\ell$ is the Jordan-Wigner transform of an arbitrary second-quantised fermionic Hamiltonian. $\textsc{Select}(H)$ is one of the main subroutines of several quantum algorithms, including state-of-the-art techniques for Hamiltonian simulation. If each term in the second-quantised Hamiltonian involves at most $k$ spin-orbitals and $k$ is a constant independent of the total number of spin-orbitals $n$ (as is the case for the majority of quantum chemistry and condensed matter models considered in the literature, for which $k$ is typically $2$ or $4$), our implementation of $\textsc{Select}(H)$ requires no ancilla qubits and uses $\mathcal{O}(n)$ Clifford+$T$ gates, with the Clifford gates applied in $\mathcal{O}(\log^2 n)$ layers and the $T$ gates in $O(\log n)$ layers. This achieves an exponential improvement in both Clifford- and $T$-depth over previous work, while maintaining linear gate count and reducing the number of ancillae to zero.

\end{abstract}

\section{Introduction}

Quantum computers have the potential to efficiently simulate quantum systems.
A particularly promising application of both near-term and fault-tolerant architectures is solving problems in quantum chemistry and materials science. 
In recent years, significant advances have been made on this front; for reviews of the major algorithmic developments, we refer the reader to Refs.~\cite{mcardle2018quantum,cao2018quantum,bauer2020quantum}.

Much of the current research in quantum simulation is concerned with the estimation of Hamiltonian spectra and preparation of energy eigenstates, which can provide insight into various properties of molecules and materials. As shown by Ref.~\cite{Abrams1999}, the quantum phase estimation algorithm~\cite{Kitaev1995quantum,Cleve1998} can be used to perform projective energy measurements, collapsing the system into a desired eigenstate with high probability if the initial state has appreciable overlap with that eigenstate. Even in the absence of a suitable initial approximation, such measurements may be applied to prepare an eigenstate by exploiting the quantum Zeno effect~\cite{Somma2008,Poulin2018}. Alternatively, approximate eigenstates may be obtained via adiabatic state preparation, given sufficient information about the gap(s) in the spectrum of the interpolating Hamiltonian~\cite{farhi2000quantum,aharonov2003adiabatic}.

Several techniques are useful for realising these schemes on a gate-based quantum computer.
For instance, the qubitisation procedure of Refs.~\cite{Low2017,Low2019} can implement the time-evolution operator $\exp(-iHt)$ or, more directly, a walk operator corresponding to $\exp[-i\arccos(Ht)]$, for a time-independent Hamiltonian $H$ and some $t \in \mathbb{R}$. Either of these operators can be used as the unitary input to phase estimation for the purpose of approximating eigenvalues and eigenstates of $H$~\cite{Abrams1999,Poulin2018,Berry2018,Babbush2018}. Adiabatic evolution can be digitally simulated by applying the truncated Dyson series algorithm of Refs.~\cite{low2018hamiltonian,Kieferova2019} for time-dependent Hamiltonian simulation, or by using the method of Ref.~\cite{wan2020fast}, which is based on quasi-adiabatic continuation~\cite{Hastings2004_LSM}. 
Approximate ground states can also be prepared using the methods of Ref.~\cite{Ge2019, lin2020}. 
All of these techniques are formulated in terms of queries to unitary oracles that encode the relevant Hamiltonian(s) in some form. One such encoding is the ``linear combination of unitaries'' (LCU) query model, motivated by the algorithms of Refs.~\cite{Childs2012,Berry2015}. In this model, the input Hamiltonian is decomposed as
\[ H = \sum_{\ell = 0}^{L-1} \alpha_\ell H_\ell, \]
where each $H_\ell$ is a time-independent unitary and the (possibly time-dependent) coefficients $\alpha_\ell$ are real and nonnegative. Information about the Hamiltonian is accessed via two oracles: $\textsc{Select}(H)$ and $\textsc{Prepare}(\alpha)$, which respectively encode the unitaries $H_\ell$ and the coefficients $\alpha_\ell$. Specifically, 
\begin{equation} \label{selHdef} \textsc{Select}(H) \coloneqq \sum_{\ell = 0}^{L-1}\ket{\ell}\bra{\ell}\otimes H_\ell \end{equation}
is a multiply-controlled operation that applies the unitary $H_\ell$ to the target register conditioned on the control register being in the state $\ket{\ell}$, and $\textsc{Prepare}(\alpha)$ is some unitary that transforms the all-zeros state of the control register as
\[ \textsc{Prepare}(\alpha): \ket{0} \mapsto \sum_{\ell=0}^{L-1}\sqrt{\frac{\alpha_\ell}{\alpha}}\ket{\ell}, \]
where $\alpha \coloneqq \sum_{\ell = 0}^{L-1}\alpha_\ell$. (In the case where the coefficients are time-dependent, $\textsc{Prepare}(\alpha)$ may be controlled on an additional register that encodes time~\cite{Kieferova2019}.)

While any operator can in principle be written as a linear combination of unitaries, some Hamiltonians are more naturally expressed in this framework. Since the complexities of the aforementioned algorithms are typically dominated by that of $\textsc{Select}(H)$ and $\textsc{Prepare}(\alpha)$,\footnote{Strictly speaking, these algorithms make calls to controlled versions of $\textsc{Select}(H)$ and $\textsc{Prepare}(\alpha)$. In our implementation, a constant number of controls can be added to $\textsc{Select}(H)$ with constant \emph{additive} overhead in the gate complexity, as will become clear in Section~\ref{sec2}.} it is important to design time- and space-efficient circuits for these oracles. The purpose of this paper is to provide an efficient construction for $\textsc{Select}(H)$ in the case where $H$ is obtained from a fermionic Hamiltonian via the Jordan-Wigner transformation~\cite{Jordan1928}, so that each $H_\ell$ is a tensor product of Pauli operators. Although our method is applicable to arbitrary fermionic Hamiltonians, it is worth noting that in many of the models considered in practice, each site interacts with only a small number of other sites. More precisely, for a fermionic Hamiltonian given in its second-quantised representation, let $k$ denote the maximum number of distinct spin-orbitals that appear in each term. For most Hamiltonians of physical interest, such as the commonly studied molecular electronic structure Hamiltonian and the Fermi-Hubbard model, $k$ does not scale with the system size. Our contribution can be stated as follows. \\

\noindent \textbf{\underline{Main result}:} For any fermionic Hamiltonian for which $k$ is a constant independent of the number of spin-orbitals $n$, we can construct a circuit for $\textsc{Select}(H)$ using zero ancilla qubits and $\mathcal{O}(n)$ Clifford and $T$ gates, with the Clifford gates performed in $\mathcal{O}(\log^2n)$ layers and the $T$ gates in $\mathcal{O}(\log n)$ layers. \\

\noindent 
This constitutes an exponential reduction in Clifford- and $T$-depth compared to existing methods. The approach of Ref.~\cite{Childs2018} can be applied to arbitrary LCU inputs but requires $\mathcal{O}(L)$ Clifford and $T$ gates and $\mathcal{O}(L)$ Clifford- and $T$-depth, and in general $L \in \mathcal{O}(n^k)$ for the type of Hamiltonians considered here. Ref.~\cite{Babbush2018} improves the gate count and depth (for both Clifford and $T$ gates) to $\mathcal{O}(n)$ for two specific $k=2$ Hamiltonians. Like Ref.~\cite{Babbush2018}, we obtain a speedup by exploiting the structure of the Jordan-Wigner encoding.\footnote{Although we focus on the Jordan-Wigner transformation in this paper, the same ideas can be used to efficiently implement \textsc{Select}($H$) for other second-quantised fermion-to-qubit mappings that have sufficient structure. In particular, our construction can be generalised to the class of mappings defined in Ref.~\cite{Havlek2017}, which includes the Jordan-Wigner and Bravyi-Kitaev~\cite{Bravyi2002,Seeley2012,Tranter2015} transformations as special cases. This would give ancilla-free circuits with the same asymptotic gate count and depth, though with larger constant factors in general~\cite{eniceicu}.} However, our circuits are completely different in structure from those in Ref.~\cite{Babbush2018}, which cannot be parallelised to sublinear-depth in any straightforward way.
Moreover, our implementation uses no ancilla qubits, in contrast to the $\sim\!\log n$ required by Refs.~\cite{Babbush2018} and~\cite{Childs2018}.

Our construction can be directly applied to asymptotically improve the circuit depth of existing fermionic simulation algorithms that are bottlenecked by $\textsc{Select}(H)$.   In Ref.~\cite{Babbush2018}, for example, the complexity of simulating the planar Fermi-Hubbard model is dominated by that of $\textsc{Select}(H)$, while $\textsc{Prepare}(\alpha)$ is extremely easy to implement as there are only three unique coefficients in the Hamiltonian. By using our circuit for $\textsc{Select}(H)$, the overall circuit depth of estimating energies via phase estimation to absolute error at most $\epsilon$ can be immediately reduced from $\widetilde{\mathcal{O}}(\alpha n/\epsilon)$ to $\widetilde{\mathcal{O}}(\alpha / \epsilon)$ in Theorem~2 of Ref.~\cite{Babbush2018} [cf.~Eq.~(27) therein], where $\widetilde{\mathcal{O}}$ hides logarithmic factors. Similarly, in Ref.~\cite{Babbush2019}, the overall depth of approximating the time-evolution operator $e^{-iHt}$ for the $k = 4$ Sachdev-Ye-Kitaev model with $n$ Majorana modes can be reduced from $\widetilde{\mathcal{O}}(n^{3.5}t)$ to $\widetilde{\mathcal{O}}(n^{2.5}t)$. 

In addition to the exponentially reduced circuit depth and minimal space overhead, an advantage of our construction lies in its simplicity and broad applicability. The circuits consist of very few different components, and take exactly the same form for all Hamiltonians with the same $k$ (though straightforward optimisations can be made if the class of input Hamiltonians is further restricted). The bulk of the gate complexity is due to a single gadget, composed entirely of controlled-\textsc{Swap} and $\textsc{cnot}$ gates. Thus, while the use of circuit depth as a complexity measure is mainly justified by long-term considerations (of prospective architectures in which many fault-tolerant gates can be executed in parallel), the simple structure of our circuits potentially makes them amenable to near-term implementation.

\section{Circuit construction} \label{sec2}

In this section, we prove our main result. We begin in subsection~\ref{sec:2.1} by developing the circuit for $\textsc{Select}(H)$ for a particular class of fermionic Hamiltonians, to illustrate the main idea. It will then become obvious how circuits for arbitrary fermionic Hamiltonians can be built, as we discuss in subsection~\ref{sec:2.3}, and that these circuits have linear gate count and polylogarithmic depth provided that $k \in \mathcal{O}(1)$. We also describe, in subsection~\ref{sec:2.2}, a simple way to substantially reduce the constant factors in the scaling of the $T$-count and $T$-depth. 

\vspace{1em}

\noindent\textit{Conventions.} Unsurprisingly, circuit diagrams are an essential part of this paper. We will use the following convention for representing operators that are controlled in a nontrivial manner on one or more qubits. Such an operator will be depicted by drawing a small solid square on the control register, connected to a box on the target register that contains the name of the operator or an abbreviation thereof. For instance, $\textsc{Select}(H)$ will be represented by
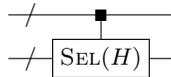
\[ {\small \Qcircuit @C=0.9em @R=1em {
&/\qw &\qw\bsquare \qwx[1] &\qw \\ 
&/\qw &\gate{\textsc{Sel}(H)} &\qw
}} \]
To clearly distinguish different registers, we will often label a control register using a computational basis state, and the target register using an arbitrary state $\ket{\psi}$. As an example, since $\ell$ is used to index the computational basis states of the control register of $\textsc{Select}(H)$ in Eq.~\eqref{selHdef}, we may add the labels $\ket{\ell}$ and $\ket{\psi}$ to the above circuit representation of $\textsc{Select}(H)$:
\[ {\small \Qcircuit @C=0.9em @R=1em {
\lstick{\ket{\ell}} &/\qw &\qw\bsquare \qwx[1] &\qw \\ 
\lstick{\ket{\psi}} &/\qw &\gate{\textsc{Sel}(H)} &\qw }
}\]
(Note that if a circuit identity holds for any computational state on the control register and any arbitrary state on the target register, it holds for all input states.) We will refer to the control register of $\textsc{Select}(H)$ as the ``selection register'' and the target register as the ``system register.''

\subsection{Main idea} \label{sec:2.1}

Our method is most easily explained by first considering quadratic fermionic Hamiltonians that consist only of terms involving two distinct spin-orbitals. The most general form of such a Hamiltonian in a second-quantised basis is
\[ \hat{H} = \sum_{p<q}\left(t_{pq}a_p^\dagger a_q + t_{pq}^* a_q^\dagger a_p + \Delta_{pq}a_p^\dagger a_q^\dagger + \Delta_{pq}^* a_q a_p\right), \]
where $a_p^\dagger$ and $a_p$ are fermionic creation and annihilation operators associated with spin-orbital $p$, and $t_{pq}, \Delta_{pq} \in \mathbb{C}$. For a system of $n$ spin-orbitals, we label the spin-orbitals from $0$ to $n-1$ in accordance with the canonical ordering chosen for the Jordan-Wigner transformation. Under this transformation, the fermionic operators are mapped to Pauli operators on $n$ qubits as $a_p \mapsto (\prod_{j=0}^{p-1} Z_j)(X_p + iY_p)/2$ and $a_p^\dagger \mapsto (\prod_{j=0}^{p-1} Z_j)(X_p - iY_p)/2$, so for $p < q$, 
\begin{align} \label{jw1} t_{pq}a_p^\dagger a_q + t_{pq}^{*}a_q^\dagger a_p &\mapsto \frac{1}{2}\vec{Z}_{p,q}\left[\mathrm{Re}(t_{pq})(X_pX_q + Y_pY_q) + \mathrm{Im}(t_{pq})(-X_pY_q + Y_pX_q)\right] \\
\label{jw2} \Delta_{pq}a^\dagger_pa^\dagger_q + \Delta_{pq}^* a_q a_p &\mapsto \frac{1}{2}\vec{Z}_{p,q}\left[\mathrm{Re}(\Delta_{pq})(X_p X_q - Y_p Y_q) + \mathrm{Im}(\Delta_{pq})(X_pY_q + Y_p X_q)\right].\end{align}
Here and throughout the paper, $Z_p$ denotes the $n$-qubit operator that acts as $Z$ on qubit $p$ and as the identity on the rest of the qubits, and similarly for $X$ and $Y$, while $\vec{Z}_{p,q} \coloneqq \prod_{j=p+1}^{q-1}Z_j$ denotes a string of $Z$ operators on all of the qubits between $p$ and $q$ (exclusive). Thus, the Jordan-Wigner transform $H$ of $\hat{H}$ is a linear combination with real coefficients of operators that all have the form $(P_1)_p \vec{Z}_{p,q} (P_2)_q$ with $P_1, P_2 \in \{X,Y\}$. Absorbing the signs of the coefficients into $P_1$, we can write 
\begin{equation*} \label{quadraticH} H = \sum_{p,q=0}^{n-1}\sum_{P_1 \in \{\pm X,\pm Y\}}\sum_{P_2 \in \{X,Y\}}\alpha_{p,q,P_1,P_2} (P_1)_p \vec{Z}_{p,q}(P_2)_q,  \end{equation*}
where the coefficients $\alpha_{p,q,P_1,P_2}$ are all nonnegative and $\alpha_{p,q,P_1,P_2} = 0$ for $p \geq q$. 

Clearly, $H$ is a linear combination of unitaries, and each of the unitaries is completely specified by the two spin-orbitals $p$ and $q$ and the Pauli operators $P_1$ and $P_2$. Accordingly, we allocate $2\ceil{\log n} + 3$ qubits
to the selection register, and encode each computational basis state $\ket{\ell}$ of the selection register as $\ket{\ell} \equiv \ket{p}\ket{q}\ket{P_1}\ket{P_2}$. The first two subregisters each contain $\ceil{\log n}$ qubits and store the binary representations of $p, q \in \{0,\dots, n-1\}$. The third and fourth subregisters, which have two qubits and one qubit, respectively, specify $P_1 \in \{\pm X, \pm Y\}$ and $P_2 \in \{X,Y\}$. By Eq.~\eqref{selHdef}, $\textsc{Select}(H)$ can then be defined by its action on computational basis states in the selection register (and an arbitrary state $\ket{\psi}$ in the system register) as
\begin{equation} \label{selH2} \textsc{Select}(H) : \ket{p}\ket{q}\ket{P_1}\ket{P_2} \otimes \ket{\psi} \mapsto \ket{p}\ket{q}\ket{P_1}\ket{P_2} \otimes (P_1)_p \vec{Z}_{p,q}(P_2)_q\ket{\psi} \end{equation}
for $p, q \in \{0, \dots, n-1\}$. (The action of $\textsc{Select}(H)$ on basis states for which $p$ and/or $q$ are out of range is unimportant, as $\textsc{Prepare}(\alpha)\ket{0}$ has no support on such states.)

To construct a circuit that implements Eq.~\eqref{selH2}, our starting point is the following circuit identity:
\begin{equation} \label{identity1} {\small \Qcircuit @C=0.5em @R=0.3em @!R {
&\qw&\qw&\qw&\qw&\qw&\qw&\qw&\targ &\qw&\targ &\qw&\qw&\qw&\qw&\qw&\qw&\qw&\qw &&&&&&&&&&\qw &\qw&\qw&\qw\\
&\qw&\qw&\qw&\qw&\qw&\qw &\targ &\ctrl{-1} &\qw &\ctrl{-1} &\targ &\qw&\qw&\qw&\qw&\qw&\qw&\qw &&&&&&&&&&\qw &\qw&\qw&\qw\\
&\qw&\qw&\qw&\qw&\qw&\targ &\ctrl{-1} &\qw&\gate{Z} &\qw&\ctrl{-1} &\targ &\qw&\qw&\qw&\qw&\qw&\qw &&&&&&&&&&\qw &\gate{Z} &\qw&\qw\\
&\qw&\qw&\qw&\qw&\targ &\ctrl{-1} &\qw&\qw&\qw&\qw&\qw&\ctrl{-1} &\targ &\qw&\qw&\qw&\qw&\qw &&&\rstick{\raisebox{-2em}{=}} &&&&&&&\qw &\gate{Z} &\qw&\qw\\
&\qw&\qw&\qw&\targ &\ctrl{-1} &\qw&\qw&\qw&\qw&\qw&\qw&\qw&\ctrl{-1} &\targ &\qw&\qw&\qw&\qw &&&&&&&&&&\qw &\gate{Z} &\qw&\qw\\
&\qw&\qw&\targ &\ctrl{-1} &\qw&\qw&\qw&\qw&\qw&\qw&\qw&\qw&\qw&\ctrl{-1} &\targ &\qw&\qw&\qw &&&&&&&&&&\qw &\gate{Z} &\qw&\qw \\
&\qw&\targ &\ctrl{-1} &\qw&\qw&\qw&\qw&\qw&\gate{Z}&\qw&\qw&\qw&\qw&\qw&\ctrl{-1} &\targ &\qw&\qw &&&&&&&&&&\qw&\qw&\qw&\qw\\
&\qw&\ctrl{-1} &\qw&\qw&\qw&\qw&\qw&\qw&\qw&\qw&\qw&\qw&\qw&\qw&\qw&\ctrl{-1} &\qw&\qw &&&&&&&&&&\qw&\qw&\qw &\qw
} }\end{equation}
which is an immediate consequence of the elementary identities
\[ {\small \Qcircuit @C=0.5em @R=0.5em {
&\qw &\gate{Z} &\targ &\qw&\qw &\rstick{\raisebox{-2em}{=}} &&&&&\qw&\targ &\gate{Z}&\qw &\qw&&&&&&&&&&&& &\qw&\qw&\targ&\qw &\qw&\rstick{\raisebox{-2em}{=}} &&&&&\qw&\targ &\qw&\qw&\qw\\
&\qw&\qw&\ctrl{-1} &\qw&\qw &&&&&&\qw&\ctrl{-1} &\gate{Z}&\qw&\qw&&&&&&&&&&&& &\qw&\gate{Z} &\ctrl{-1} &\qw&\qw&&&&&&\qw&\ctrl{-1} &\gate{Z} &\qw&\qw
}} \]
and the fact that \textsc{cnot} is self-inverse. 
The analogue of Eq.~\eqref{identity1} for an arbitrary number of qubits and with the two $Z$ operators on a different pair of qubits is obvious. Letting $Q_{P_1}$ denote the Pauli operator such that $Q_{P_1}Z = iP_1$ for $P_1 \in \{\pm X,\pm Y\}$ (i.e., $Q_{\pm X} = \pm Y$ and $Q_{\pm Y} = \mp X$), it follows that 
\begin{equation} \label{identity2} {\small \Qcircuit @C=0.5em @R=0.1em @!R {
&&\qw&\qw&\qw&\qw&\qw&\qw&\qw&\targ &\qw&\targ &\qw&\qw&\qw&\qw&\qw&\qw&\qw&\qw&\qw&\qw &&&&&&&&&&&\qw &\qw&\qw&\qw\\
&&\qw&\qw&\qw&\qw&\qw&\qw &\targ &\ctrl{-1} &\qw &\ctrl{-1} &\targ &\qw&\qw&\qw&\qw&\qw&\qw&\qw&\qw&\qw &&&&&&&&&&&\qw &\qw&\qw&\qw\\
&\lstick{p} &\qw&\qw&\qw&\qw&\qw&\targ &\ctrl{-1} &\qw&\gate{Z} &\qw&\ctrl{-1} &\targ &\qw&\qw&\qw&\qw&\gate{-i} &\gate{Q_{P_1}} &\qw &\qw &&&&&&&&&&\lstick{p} &\qw &\gate{P_1} &\qw&\qw\\
&&\qw&\qw&\qw&\qw&\targ &\ctrl{-1} &\qw&\qw&\qw&\qw&\qw&\ctrl{-1} &\targ &\qw&\qw&\qw&\qw&\qw&\qw&\qw &&&\rstick{\raisebox{-2em}{=}} &&&&&&&&\qw &\gate{Z} &\qw&\qw\\
&&\qw&\qw&\qw&\targ &\ctrl{-1} &\qw&\qw&\qw&\qw&\qw&\qw&\qw&\ctrl{-1} &\targ &\qw&\qw&\qw&\qw&\qw&\qw &&&&&&&&&&&\qw &\gate{Z} &\qw&\qw\\
&&\qw&\qw&\targ &\ctrl{-1} &\qw&\qw&\qw&\qw&\qw&\qw&\qw&\qw&\qw&\ctrl{-1} &\targ &\qw&\qw&\qw&\qw&\qw &&&&&&&&&&&\qw &\gate{Z} &\qw&\qw \\
&\lstick{q} &\qw&\targ &\ctrl{-1} &\qw&\qw&\qw&\qw&\qw&\gate{Z}&\qw&\qw&\qw&\qw&\qw&\ctrl{-1} &\targ &\qw &\gate{P_2}&\qw&\qw &&&&&&&&&&\lstick{q}&\qw&\gate{P_2}&\qw&\qw\\
&&\qw&\ctrl{-1} &\qw&\qw&\qw&\qw&\qw&\qw&\qw&\qw&\qw&\qw&\qw&\qw&\qw&\ctrl{-1} &\qw&\qw&\qw&\qw &&&&&&&&&&&\qw&\qw&\qw &\qw
} }\end{equation}
Therefore, if the $n$ qubits in the system register are ordered such that the qubit corresponding to spin-orbital~$0$ is on the top wire and the qubit corresponding to spin-orbital $n-1$ is on the bottom, the circuit on the left-hand side would implement the term $(P_1)_p \vec{Z}_{p,q}(P_2)_q\ket{\psi}$ for a particular $p, q, P_1, P_2$. 

From here, we would obtain a circuit for $\textsc{Select}(H)$ if we were to control the $Q_{P_1}$, $P_2$, and $Z$ operators in the circuit of Eq.~\eqref{identity2} on the selection register such that
\begin{enumerate}
\item the states $\ket{P_1}$ and $\ket{P_2}$ of the third and fourth selection subregisters determine which Pauli operators $Q_{P_1}$ and $P_2$ represent, and
\item conditioned on the first two selection subregisters being in the state $\ket{p}\ket{q}$, $Q_{P_1}$ and one of the $Z$ operators are applied to qubit $p$ of the system register, while $P_2$ and the other $Z$ operator are applied to qubit $q$. 
\end{enumerate}

Condition (1) can be straightforwardly satisfied by constructing $\textsc{Select}(Q)$ and $\textsc{Select}(P)$ operators that choose the appropriate $Q_{P_1}$ and $P_2$ according to the states $\ket{P_1}$ and $\ket{P_2}$. For concreteness, suppose that $\ket{P_1} = \ket{00}, \ket{01}, \ket{10}, \ket{11}$ for $P_1 = X, -X, Y, -Y$, respectively, and $\ket{P_2} = \ket{0}, \ket{1}$ for $P_2 = X, Y$, respectively. Then, $\textsc{Select}(Q)$ and $\textsc{Select}(P)$ can be implemented as
\begin{equation} \label{SelQP} {\small \Qcircuit @C=0.9em @R=0.5em {
&/\qw &\qw \bsquare \qwx[2] &\qw &&&&&\gate{Z} &\ctrlo{2} &\ctrl{2} &\qw &&&&&&& &\qw \bsquare \qwx[2] &\qw &&&& &\ctrlo{2} &\ctrl{2} &\qw \\
&&&&\rstick{=} &&&&\gate{Z} &\qw &\qw &\qw &&&&&&&&&&\rstick{=} &&&&& \\
&\qw &\gate{\textsc{Sel}(Q)} &\qw &&&&&\qw &\gate{Y} &\gate{X} &\qw &&&&&&& &\gate{\textsc{Sel}(P)} &\qw &&&&&\gate{X} &\gate{Y} &\qw
}} \end{equation}

Condition (2) requires the ability to target a particular qubit in the system register depending on the states of the selection subregisters that encode $\ket{p}$ and $\ket{q}$. For this purpose, we define for any single-qubit unitary $U$ a $(\ceil{\log n} + n)$-qubit operator $\textsc{Inject}(U)$. When applied to $\ket{x}\ket{\psi}$, where $x \in \{0,\dots, n-1\}$ (encoded in binary) and $\ket{\psi}$ is an arbitrary $n$-qubit state, $\textsc{Inject}(U)$ implements $U$ on qubit $x$ of $\ket{\psi}$ and acts as the identity on the other qubits, i.e.,
\[ \textsc{Inject}(U) : \ket{x}\ket{\psi} \mapsto \ket{x} \left[\left(I^{\otimes x} \otimes U \otimes I^{\otimes (n-x-1)}\right)\ket{\psi} \right]. \]
To synthesise $\textsc{Inject}(U)$ for any $U$, we use a $(\ceil{\log n} + n)$-qubit operator $\textsc{SwapUp}$, defined as follows. For any $x \in \{0,\dots, n-1\}$ and $n$-qubit product state $\bigotimes_{y=0}^{n-1}\ket{\varphi_y}_y$,
\begin{equation} \label{swapup} \textsc{SwapUp} : \ket{x} \bigotimes_{y=0}^{n-1} \ket{\varphi_y}_y \mapsto \ket{x}\ket{\varphi_x}_0 \bigotimes_{y=1}^{n-1}\ket{\varphi_{\sigma(y)}}_y, \end{equation}
where $\sigma$ is a fixed\footnote{In principle, any permutation $\sigma$ such that $\sigma(0) = x$ would work---the action of $\sigma$ on $\{1,\dots, n-1\}$ is not relevant to the correctness of our construction. However, the implementation of \textsc{SwapUp} in Appendix~\ref{appendix1} yields a particular permutation $\sigma$ (for each $n$).} permutation of $\{0,\dots, n-1\}$ such that $\sigma(0) = x$. In other words, conditioned on the $\ceil{\log n}$-qubit control register being in the state $x$ for $x \in \{0,\dots, n-1\}$, $\textsc{SwapUp}$ moves the state $\ket{\varphi_x}$ of the qubit indexed by $x$ in the target register up to the first qubit of the target register, and permutes the states of the other qubits in some way. Ref.~\cite{low2018trading} shows that $\textsc{SwapUp}$ can be implemented without ancilla qubits using $\mathcal{O}(n)$ Clifford and $T$ gates, $O(\log^2 n)$ Clifford-depth, and $O(\log n)$ $T$-depth. We sketch the construction in Appendix~\ref{appendix1}. 
From Eq.~\eqref{swapup}, it is easy to see that $\textsc{Inject}(U) = \textsc{SwapUp}^\dagger (U \otimes I^{(n-1)})\textsc{SwapUp}$. \textsc{SwapUp} permutes the qubits in the target register in such a way that the state of qubit $x$ is moved up to the first qubit, $U$ is then applied to the first qubit, and $\textsc{SwapUp}^\dagger$ undoes the permutation. 
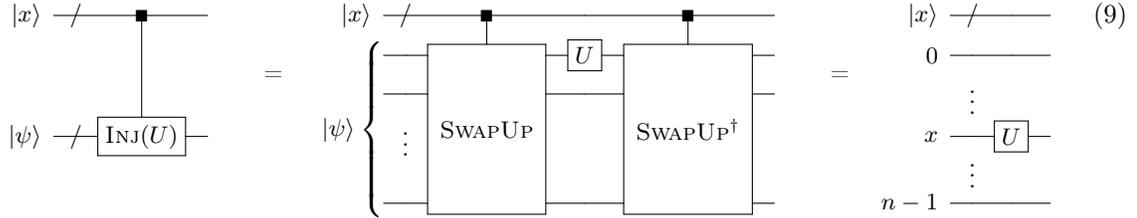
\begin{equation} \label{InjU} {\small \Qcircuit @C=0.9em @R=0.5em {
&\lstick{\ket{x}} &/\qw &\qw \bsquare \qwx[4] &\qw &&&&&&&&\lstick{\ket{x}} &/\qw &\qw \bsquare \qwx[2] &\qw &\qw \bsquare\qwx[2] &\qw &&&&&&&&\lstick{\ket{x}} &/\qw &\qw &\qw \\ &&&&&&&&&&&&&&&&&&&&&\\
&&&&&&\rstick{\raisebox{-2em}{=}}&&&&&&&\qw &\multigate{5}{\textsc{SwapUp}} &\gate{U} &\multigate{5}{\textsc{SwapUp}^\dagger} &\qw &&\rstick{\raisebox{-2em}{=}}&&&&&&\lstick{0} &\qw &\qw &\qw \\
&&&&&&&&&&&&&\qw &\ghost{\textsc{SwapUp}} &\qw &\ghost{\textsc{SwapUp}^\dagger} &\qw &&&&&&&&&\vdots \\
&\lstick{\ket{\psi}} &/\qw &\gate{\textsc{Inj}(U)} &\qw &&&&&&&&\lstick{\raisebox{0.8em}{$\ket{\psi}\enspace\,$}} &\vdots &&&&&&&&&&& &\lstick{x} &\qw &\gate{U} &\qw \\
&&&&&&&&&&&& & & &&&&&&&&&&&&\vdots\\ &&&&&&&&&&&&&&&&&&&&&\\
&&&&&&&&&&&&&\qw &\ghost{\textsc{SwapUp}} &\qw &\ghost{\textsc{SwapUp}^\dagger} &\qw &&&&&&&&\lstick{n-1} &\qw &\qw &\qw
\gategroup{3}{13}{8}{13}{1em}{\{}
}}
\end{equation}
(The third circuit above illustrates the effect of $\textsc{Inject}(U)$ in the special case where the input to the control register is a computational basis state, whereas the second circuit implements $\textsc{Inject}(U)$ on arbitrary inputs.)

Hence, we can ensure that the two $Z$ operators in Eq.~\eqref{identity2} are applied to qubits $p$ and $q$ of the system register conditioned on the state of the first two selection subregisters being $\ket{p}\ket{q}$ by implementing two $\textsc{Inject}(Z)$ operators, one with $\ket{p}$ as the control register and the other with $\ket{q}$ as the control register.
To correctly apply $Q_{P_1}$ and $P_2$, we use the $\textsc{Select}(Q)$ and $\textsc{Select}(P)$ circuits constructed in Eq.~\eqref{SelQP} in conjunction with $\textsc{SwapUp}$ to form $\textsc{Inject-Select}(Q)$ and $\textsc{Inject-Select}(P)$. This is shown below for $\textsc{Inject-Select}(Q)$; the construction of $\textsc{Inject-Select}(P)$ is analogous. 
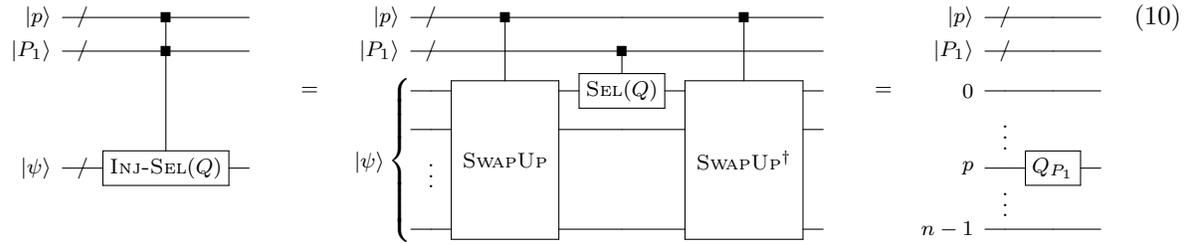
\begin{equation} \label{InjSelQ} {\footnotesize \Qcircuit @C=0.9em @R=0.5em {
&\lstick{\ket{p}} &/\qw &\qw \bsquare \qwx[7] &\qw &&&&&&&&\lstick{\ket{p}} &/\qw &\qw \bsquare \qwx[5] &\qw &\qw \bsquare\qwx[5] &\qw &&&&&&&&\lstick{\ket{p}} &/\qw &\qw &\qw \\
&&&&&&&&&&&&&&&&&&&&&&& \\
&&&&&&&&&&&&&&&&&&&&&&& \\
&\lstick{\ket{P_1}} &/\qw &\qw\bsquare &\qw &&&&&&&&\lstick{\ket{P_1}} &/\qw &\qw &\qw\bsquare \qwx[2]&\qw &\qw &&&&&&&&\lstick{\ket{P_1}} &/\qw &\qw &\qw\\
&&&&&&&&&&&&&&&&&&&&&&& \\
&&&&&&\rstick{=}&&&&&&&\qw &\multigate{5}{\textsc{SwapUp}} &\gate{\textsc{Sel}(Q)} &\multigate{5}{\textsc{SwapUp}^\dagger} &\qw &&\rstick{=}&&&&&&\lstick{0} &\qw &\qw &\qw \\
&&&&&&&&&&&&&\qw &\ghost{\textsc{SwapUp}} &\qw &\ghost{\textsc{SwapUp}^\dagger} &\qw && &&&&&&&\vdots \\
&\lstick{\ket{\psi}} &/\qw &\gate{\textsc{Inj-Sel}(Q)} &\qw &&&&&&&&\lstick{\raisebox{0.8em}{$\ket{\psi}\enspace\,$}} &\vdots &&&&&&&&&&& &\lstick{p} &\qw &\gate{Q_{P_1}} &\qw \\
&&&&&&&&&&&& & & &&&&&&&&&&&&\vdots\\ &&&&&&&&&&&&&&&&&&&&&\\
&&&&&&&&&&&&&\qw &\ghost{\textsc{SwapUp}} &\qw &\ghost{\textsc{SwapUp}^\dagger} &\qw &&&&&&&&\lstick{n-1} &\qw &\qw &\qw
\gategroup{6}{13}{11}{13}{1em}{\{}
}}
\end{equation}

With these components in hand, we can assemble the circuit for $\textsc{Select}(H)$:
\begin{equation} \label{selHcircuit0} {\footnotesize \Qcircuit @C=0.9em @R=0.5em {
&\lstick{\ket{\ell}} &/\qw &\qw \bsquare \qwx[11] &\qw &&&&&&&&&&\lstick{\ket{p}} &/\qw &\qw &\qw\bsquare\qwx[11] &\qw &\qw &\qw &\qw\bsquare\qwx[11] &\qw &\qw \\
&&&&&&&&&&&&&&&&\\&&&&&&&&&&&&&&&& \\
&&&&&&&&&&&&&&\lstick{\ket{q}} &/\qw &\qw &\qw &\qw\bsquare\qwx[8] &\qw &\qw &\qw &\qw\bsquare\qwx[8] &\qw \\
&&&&&&&&&&&&&&&&\\&&&&&&&&&&&&&&&&\\
&&&&&&\rstick{\raisebox{-2em}{=}} &&&&&&\lstick{\raisebox{1.75em}{$\ket{\ell}$\enspace\,\,}} &&\lstick{\ket{P_1}} &/\qw &\qw &\qw &\qw &\qw &\qw &\qw\bsquare &\qw &\qw \\
&&&&&&&&&&&&&&&&\\&&&&&&&&&&&&&&&&\\
&&&&&& &&&&&&&&\lstick{\ket{P_2}} &\qw &\qw &\qw &\qw &\qw &\qw &\qw &\qw\bsquare &\qw\\
&&&&&&&&&&&&&&&&\\
&\lstick{\ket{\psi}} &/\qw &\gate{\textsc{Sel}(H)} &\qw &&&&&&&&&&\lstick{\ket{\psi}} &/\qw &\gate{\rotatebox{90}{\textsc{Ladder}}} &\gate{\textsc{Inj}(Z)} &\gate{\textsc{Inj}(Z)} &\gate{\rotatebox{90}{$\textsc{Ladder}^\dagger$}} &\gate{-i} &\gate{\textsc{Inj-Sel}(Q)} &\gate{\textsc{Inj-Sel}(P)} &\qw
\gategroup{1}{13}{10}{13}{1.3em}{\{}} }
\end{equation}
where \textsc{Ladder} denotes the operator implemented by the ladder-like sequence of $n-1$ $\textsc{cnot}$ gates in Eq.~\eqref{identity2}:
\begin{equation} \label{ladder} {\small \Qcircuit @C=0.5em @R=0.5em{ &\qw &\multigate{4}{\rotatebox{90}{\textsc{Ladder}}} &\qw &\qw &&&&&&&&\qw&\qw&\qw&\qw&\targ &\qw&\qw\\
&\qw &\ghost{\rotatebox{90}{\textsc{Ladder}}} &\qw &\qw &&&&&&&&\qw&\qw&\qw&\targ &\ctrl{-1}&\qw&\qw \\
&\qw &\ghost{\rotatebox{90}{\textsc{Ladder}}} &\qw &\qw &&\rstick{\coloneqq} &&&&&&\qw&\qw &\targ &\ctrl{-1} &\qw&\qw&\qw \\
&\qw &\ghost{\rotatebox{90}{\textsc{Ladder}}} &\qw &\qw &&&&&&&&\qw&\targ &\ctrl{-1} &\qw&\qw&\qw&\qw\\
&\qw &\ghost{\rotatebox{90}{\textsc{Ladder}}} &\qw &\qw &&&&&&&&\qw&\ctrl{-1} &\qw&\qw&\qw&\qw&\qw \\
}}\end{equation}
By comparing the circuit in Eq.~\eqref{selHcircuit0} to that in Eq.~\eqref{identity2}, it can be verified that the former correctly implements $\textsc{Select}(H)$ as it is defined in Eq.~\eqref{selH2}. When the selection register is in the computational basis state $\ket{p}\ket{q}\ket{P_1}\ket{P_2}$, the two $\textsc{Inject}(Z)$ operators apply $Z$ operators on qubits $p$ and $q$ in the system register, between the two sequences of $\textsc{cnot}$ gates corresponding to $\textsc{Ladder}$ and $\textsc{Ladder}^\dagger$. Then, $\textsc{Inject-Select}(Q)$ applies $Q_{P_1}$ to qubit $q$ and $\textsc{Inject-Select}(P)$ applies $P_2$ to qubit $q$. Thus, by Eq.~\eqref{identity2}, the circuit applies the operator $(P_1)_p \vec{Z}_{p,q}(P_2)_q$ on the system register conditioned on the state of the selection register being $\ket{p}\ket{q}\ket{P_1}\ket{P_2}$, as required by Eq.~\eqref{selH2}. This holds for all of the computational basis states, and therefore for arbitrary states of the selection register. 

It is clear from Eqs.~\eqref{SelQP}, \eqref{InjU}, and~\eqref{InjSelQ} that the only non-Clifford gates in the circuit of Eq.~\eqref{selHcircuit0} are the \textsc{SwapUp} gadgets used to construct the  ``\textsc{Inject}'' operators. As shown in Appendix~\ref{appendix1}, \textsc{SwapUp} on $\ceil{\log n} + n$ qubits can be implemented with $\mathcal{O}(n)$ $T$ gates and $\mathcal{O}(\log n)$ $T$-depth. Each of $\textsc{Inject}(Z)$, $\textsc{Inject-Select}(Q)$, and $\textsc{Inject-Select}(P)$ uses one $\textsc{SwapUp}$ and one $\textsc{SwapUp}^\dagger$. Since the number of these operators is a constant independent of $n$, the total $T$-count of the circuit in Eq.~\eqref{selHcircuit0} is $\mathcal{O}(n)$ and the total $T$-depth is $\mathcal{O}(\log n)$. The exact $T$ costs of the components are provided by Table~\ref{table:costs}.

\begin{table}
\begin{centering}
\begingroup
\renewcommand{\arraystretch}{1.2}
\begin{tabular}{  c | c | c | c}
    circuit component & $T$-count & $T$-depth & \makecell{\# of elementary gates \\ to control} \\ \hline
    $\textsc{Inject}(Z)$ &$28(n-1)$ &$32\ceil{\log n}$ &\multirow{2}{*}{$1$}\\ \cline{1-3}
    $\textsc{Inject}^*(Z)$ &$8(n-1)$ &$8\ceil{\log n}$ \\ \hline
    $\textsc{Inject-Select}(Q)$ &$28(n-1)$ &$32\ceil{\log n}$ &\multirow{2}{*}{$4$} \\ \cline{1-3}
    $\textsc{Inject-Select}^*(Q)$ &$16(n-1)$ &$16\ceil{\log n}$ \\ \hline
    $\textsc{Inject-Select}(P)$ &$28(n-1)$ &$32\ceil{\log n}$ &\multirow{2}{*}{$2$} \\ \cline{1-3}
    $\textsc{Inject-Select}^*(P)$ &$16(n-1)$ &$16\ceil{\log n}$
\end{tabular}
\caption{$T$-count and $T$-depth of each of the main components used to construct circuits for $\textsc{Select}(H)$ in Section~\ref{sec2}. (The Clifford-count and Clifford-depth are $\mathcal{O}(n)$ and $\mathcal{O}(\log^2 n)$, respectively, for all of the components.) The fourth column specifies the number of one- or two-qubit (Clifford) gates to which controls need to be added in order to construct the controlled version of the operator in the first column.} \label{table:costs}
\endgroup
\end{centering}
\end{table}

The Clifford complexity of $\textsc{Inject}(Z)$, $\textsc{Inject-Select}(Q)$, and $\textsc{Inject-Select}(P)$ is dominated by that of $\textsc{SwapUp}$ and its inverse, which have Clifford-count $\mathcal{O}(n)$ and Clifford-depth $\mathcal{O}(\log^2 n)$ [cf.~Appendix~\ref{appendix1}]. The circuit for the \textsc{Ladder} operator given in Eq.~\eqref{ladder} has Clifford-depth $n-1$; however, as shown in Appendix~\ref{appendix3}, the same operator can be implemented using an ancilla-free circuit consisting of $\mathcal{O}(n)$ \textsc{cnot}s arranged in $\mathcal{O}(\log n)$ layers. It follows that the total Clifford-count of the circuit in Eq.~\eqref{selHcircuit0} is $\mathcal{O}(n)$ and the total Clifford-depth is $\mathcal{O}(\log^2n)$.

Although the circuit of Eq.~\eqref{selHcircuit0} achieves $\mathcal{O}(n)$ $T$-count and $\mathcal{O}(\log n)$ $T$-depth, we can reduce the constant factors hidden under the big $\mathcal{O}$ if desired by making a few modifications, as demonstrated in the following subsection. We conclude this subsection by noting that the controlled version of $\textsc{Select}(H)$ can be constructed by adding controls to a very small number of gates in the circuit---namely, the $Z$ operator in each $\textsc{Inject}(Z)$ [cf.~Eq.~\eqref{InjU}], the Pauli and controlled-Pauli operators in $\textsc{Inject-Select}(Q)$ and $\textsc{Inject-Select}(P)$ [cf.~Eqs.~\eqref{SelQP} and~\eqref{InjSelQ}], and the $(-i)$-phase gate (by implementing an $S^\dagger$ gate on the control qubit). When these operators are not applied, the circuit implements the identity since the \textsc{Ladder} and \textsc{SwapUp} operators are cancelled by their inverses. Therefore, controlling the entire circuit on any constant number of qubits incurs only constant additive gate complexity (which can be quantified using the fourth column of Table~\ref{table:costs}). This is important because algorithms that use the LCU query model generally require access to controlled-$\textsc{Select}(H)$. 


\subsection{Reducing the constant factors} \label{sec:2.2}

Before generalising the $\textsc{Select}(H)$ circuit in subsection~\ref{sec:2.1} to arbitrary fermionic Hamiltonians, we provide more efficient versions of the main circuit components, which can be used to reduce the $T$-count and $T$-depth by constant multiplicative factors. The Clifford-count and Clifford-depth are reduced by constant factors as well. However, we focus on the $T$ complexity in this subsection because $T$ gates are the bottleneck in many models of fault-tolerant quantum computation, notably those that are based on topological error correcting codes. In these settings, $T$ gates require significantly more time and physical qubits to implement than Clifford gates~\cite{fowler2012bridge}, and even a constant-factor improvement in the $T$ complexity may be useful.

The strategy is to replace all of the \textsc{SwapUp} operators by a particular phase-incorrect \textsc{SwapUp} operator, which is based on a phase-incorrect \textsc{Toffoli} gate introduced in Ref.~\cite{Barenco1995}. As pointed out by Ref.~\cite{low2018trading}, this phase-incorrect \textsc{SwapUp}, which we will call $\textsc{SwapUp}^*$, can be implemented using $4(n-1)$ $T$ gates that are applied in $4\ceil{\log n}$ layers, a considerable reduction from the $14(n-1)$ $T$-count and $16\ceil{\log n}$ $T$-depth of \textsc{SwapUp}. The circuit for $\textsc{SwapUp}^*$ is described in Appendix~\ref{appendix2}. In the computational basis, $\textsc{SwapUp}^*$ has the same matrix elements as $\textsc{SwapUp}$ up to sign, i.e., for any computational basis state $\ket{z}$, 
\[ \textsc{SwapUp}^*\ket{z} = \pm \textsc{SwapUp}\ket{z} = \pm \ket{z'}, \]
where $\ket{z'} = \textsc{SwapUp}\ket{z}$ is also a computational basis state. Hence, we can write $\textsc{SwapUp}^* = D\cdot \textsc{SwapUp}$ for some operator $D$ that is diagonal in the computational basis, with eigenvalues $\pm 1$. This implies that $\textsc{Inject}(Z)$ would still be implemented correctly if \textsc{SwapUp} and $\textsc{SwapUp}^\dagger$ in the circuit of Eq.~\eqref{InjU} were replaced with $\textsc{SwapUp}^*$ and $\textsc{SwapUp}^{*\dagger}$:
\begin{align*} \textsc{SwapUp}^{*\dagger} Z_1 \textsc{SwapUp}^* &= (\textsc{SwapUp}^\dagger D^\dagger)Z_1(D \cdot \textsc{SwapUp}) \\ &= \textsc{SwapUp}^\dagger Z_1 \textsc{SwapUp}  = \textsc{Inject}(Z),
\end{align*}
where the second equality follows from the fact that $D$ commutes with $Z_1$ and is unitary. We denote this implementation of $\textsc{Inject}(Z)$ by $\textsc{Inject}^*(Z)$.\footnote{We clarify that unlike in the case of \textsc{SwapUp} and $\textsc{SwapUp}^*$, which are different operators, $\textsc{Inject}(Z)$ and $\textsc{Inject}^*(Z)$ designate different circuit implementations of the \emph{same} operator. The same goes for $\textsc{Inject-Select}(Q)$ and $\textsc{Inject-Select}^*(Q)$, and $\textsc{Inject-Select}(P)$ and $\textsc{Inject-Select}^*(P)$.}  By the same token, $\textsc{SwapUp}^*$ can be used to construct $\textsc{Inject}(U)$ for any $U$ that is diagonal in the computational basis. 

On the other hand, the circuit in Eq.~\eqref{InjSelQ} would not correctly implement $\textsc{Inject-Select}(Q)$ if $\textsc{SwapUp}^*$ were used instead of $\textsc{SwapUp}$, since $X$ and $Y$ are not diagonal in the computational basis. To reduce the $T$ cost of $\textsc{Inject-Select}(Q)$, we modify the circuits so that $X$ and $Y$ are ``injected'' separately by applying $\textsc{Inject}^*(Z)$ conjugated by basis change operators, as follows:
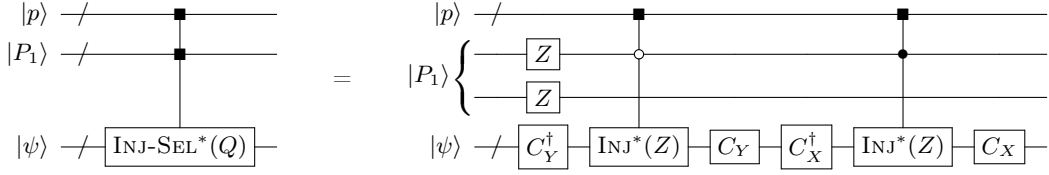
\[ {\small \Qcircuit @C=0.9em @R=0.5em {
&\lstick{\ket{p}} &/\qw &\qw\bsquare\qwx[2] &\qw &&&&&&&& &\lstick{\ket{p}} &/\qw &\qw &\qw\bsquare\qwx[2] &\qw &\qw &\qw\bsquare\qwx[2] &\qw &\qw \\
\\
&\lstick{\ket{P_1}} &/\qw &\qw\bsquare \qwx[2] &\qw &&\rstick{\raisebox{-2.55em}{=}}&&&&&& &\lstick{\raisebox{-2.55em}{$\ket{P_1} \enspace$}} &\qw &\gate{Z} &\ctrlo{2} &\qw &\qw &\ctrl{2} &\qw &\qw \\
&&&&&&&&&&&& &&\qw &\gate{Z} &\qw &\qw &\qw &\qw &\qw &\qw \\
&\lstick{\ket{\psi}} &/\qw &\gate{\textsc{Inj-Sel}^*(Q)} &\qw &&&&&&&& &\lstick{\ket{\psi}} &/\qw &\gate{C_Y^{\dagger}}&\gate{\textsc{Inj}^*(Z)} &\gate{{C_Y}} &\gate{C_X^{\dagger}}&\gate{\textsc{Inj}^*(Z)} &\gate{{C_X}} &\qw 
\gategroup{3}{14}{4}{14}{1em}{\{}
}}
\]
using $C_X$ and $C_Y$ to represent $B_X^{\otimes n}$ and $B_X^{\otimes n}$, where $B_X$ and $B_Y$ are Clifford gates for which $B_X Z B_X^\dagger = X$ and $B_Y Z B_Y^\dagger = Y$. Note that controlled-$\textsc{Inject}(Z)$ can be constructed by simply adding a control to the $Z$ operator in $\textsc{Inject}^{*}(Z)$. The construction for $\textsc{Inject-Select}(Q)$ is similar [cf.~Eq.~\eqref{SelQP}]. We denote these alternative implementations of $\textsc{Inject-Select}(Q)$ and $\textsc{Inject-Select}(P)$ by $\textsc{Inject-Select}^*(Q)$ and $\textsc{Inject-Select}^*(P)$.

The $T$-count and $T$-depth of each of these improved circuit components follow directly from the $T$ cost of $\textsc{SwapUp}^{*}$, and are listed in Table~\ref{table:costs}. $\textsc{Inject}(Z)$, $\textsc{Inject-Select}(P)$, and $\textsc{Inject-Select}(Q)$ can always be replaced with their asterisked counterparts to minimise the complexity. For example, the circuit in Eq.~\eqref{selHcircuit0}, which implements $\textsc{Select}(H)$ for quadratic fermionic Hamiltonians, has $T$-count $112(n-1)$ and $T$-depth $128\ceil{\log n}$. Replacing the components in Eq.~\eqref{selHcircuit0} by their improved versions would reduce the total $T$-count to $48(n-1)$ and the $T$-depth to $48\ceil{\log n}$.\footnote{As a side note, the \textsc{Toffoli}-count of any of the non-asterisked components can be obtained by dividing the corresponding $T$-count in Table~\ref{table:costs} by $7$, and the \textsc{Toffoli}-depth can be obtained by dividing the $T$-depth by $4$. The asterisked operators are not constructed using \textsc{Toffoli}s. See Appendix~\ref{appendix1} and~\ref{appendix2} for details.}

\subsection{Generalising to arbitrary $k$} \label{sec:2.3}

We can extend the ideas of subsections~\ref{sec:2.1} and~\ref{sec:2.2} to devise ancilla-free implementations of $\textsc{Select}(H)$ for the Jordan-Wigner transforms of arbitrary fermionic Hamiltonians. If at most $k$ distinct spin-orbitals are involved in each term of the fermionic Hamiltonian, the resulting circuit has Clifford- and $T$-count $\mathcal{O}(kn)$, Clifford-depth $\mathcal{O}(k\log^2n)$, and $T$-depth $\mathcal{O}(k\log n)$. To help illustrate the concepts by way of circuit diagrams, we will use $k=4$ Hamiltonians as a concrete example. 

In its second-quantised representation, each term in a general fermionic Hamiltonian is a product of interaction operators $a_p^\dagger a_q$ or $a_p^\dagger a_q^\dagger$ (with $p < q$) and their Hermitian conjugates, and number operators $n_p \coloneqq a_p^\dagger a_p$.\footnote{In theory, the Hamiltonian could contain terms that are linear, cubic, etc. in the fermionic operators. Our method can be used to implement these terms as well (basically, by exploiting the identity in Eq.~\eqref{identity1} except with an odd number of $Z$ operators), but since they rarely appear in Hamiltonians of interest, we omit them for simplicity.} By definition of $k$, Hamiltonians with $k = 4$ may include such terms as $a_p^\dagger a_q a_r^\dagger a_s + \mathrm{h. c.}$, $a_p^\dagger a_q n_r n_s+ \mathrm{h. c.}$, $n_p$, and $n_p n_q n_r$, to list a few examples. 
The circuit for $\textsc{Select}(H)$ can be constructed in two main parts. Loosely speaking, one part of the circuit implements interaction operators and the other part implements number operators. 

As we saw in subsection~\ref{sec:2.1}, an interaction operator involving two spin-orbitals $p$ and $q$ is mapped to linear combinations of Pauli ``strings'' of the form $(P_1)_{p}\vec{Z}_{p,q}(P_2)_q$ under the Jordan-Wigner transformation , with $P_1,P_2 \in \{X, Y\}$ [cf.~Eqs.~\eqref{jw1} and~\eqref{jw2}]. More generally, any product of interaction operators is mapped to a linear combination of products of such Pauli strings. For instance, $a_p^\dagger a_q a_r^\dagger a_s + \mathrm{h. c.}$ (for $p < q < r <s$) becomes a linear combination of $(P_1)_p\vec{Z}_{p,q}(P_2)_q (P_3)_r\vec{Z}_{r,s}(P_4)_q$, for $P_1,P_2,P_3,P_4 \in \{X,Y\}$. Hence, the circuit in Eq.~\eqref{selHcircuit0}, which implements $\textsc{Select}(H)$ in the special case that the Hamiltonian consists only of interactions between two spin orbitals, can be easily expanded to implement Hamiltonians containing arbitrary products of interaction operators. The key observation is that the identities in Eqs.~\eqref{identity1} and~\eqref{identity2} hold analogously for any number of pairs of $Z$ operators, e.g., for two pairs of $Z$ operators, we have
\begin{equation} \label{identity4} {\small \Qcircuit @C=0.5em @R=0.1em  @!R{
&&\qw&\qw&\qw&\qw&\qw&\qw&\qw&\qw&\qw &\targ &\qw &\targ &\qw&\qw&\qw&\qw&\qw&\qw&\qw&\qw&\qw&\qw&\qw&\qw &&&&&&&&&&&\qw &\qw&\qw&\qw\\
&\lstick{p} &\qw&\qw&\qw&\qw&\qw&\qw&\qw&\qw&\targ &\ctrl{-1} &\gate{Z} &\ctrl{-1} &\targ &\qw&\qw&\qw&\qw&\qw&\qw&\qw&\gate{-i} &\gate{Q_{P_1}} &\qw &\qw &&&&&&&&&&\lstick{p} &\qw &\gate{P_1} &\qw&\qw\\
&&\qw&\qw&\qw&\qw&\qw&\qw&\qw&\targ &\ctrl{-1} &\qw&\qw&\qw&\ctrl{-1} &\targ &\qw&\qw&\qw&\qw&\qw&\qw&\qw &\qw&\qw&\qw &&& &&&&&&&&\qw &\gate{Z} &\qw&\qw\\
&&\qw&\qw&\qw&\qw&\qw&\qw&\targ &\ctrl{-1} &\qw&\qw&\qw&\qw&\qw&\ctrl{-1} &\targ &\qw&\qw&\qw&\qw&\qw&\qw&\qw&\qw&\qw &&&&&&&&&&&\qw &\gate{Z} &\qw&\qw\\
&\lstick{q} &\qw&\qw &\qw&\qw&\qw&\targ &\ctrl{-1} &\qw&\qw&\qw&\gate{Z}&\qw&\qw&\qw&\ctrl{-1} &\targ &\qw&\qw&\qw&\qw &\qw&\gate{P_2}&\qw&\qw &&&\rstick{\raisebox{-2em}{=}}&&&&&&&\lstick{q}&\qw &\gate{P_2} &\qw&\qw \\
&&\qw&\qw&\qw&\qw&\targ &\ctrl{-1} &\qw&\qw&\qw&\qw&\qw&\qw&\qw&\qw&\qw&\ctrl{-1} &\targ &\qw &\qw&\qw&\qw &\qw&\qw&\qw&&&&&&&&&&&\qw&\qw&\qw&\qw\\
&\lstick{r} &\qw&\qw&\qw&\targ &\ctrl{-1} &\qw&\qw&\qw&\qw&\qw&\gate{Z}&\qw&\qw&\qw&\qw&\qw&\ctrl{-1}&\targ &\qw&\qw&\gate{-i} &\gate{Q_{P_3}} &\qw &\qw &&&&&&&&&&\lstick{r}&\qw&\gate{P_3}&\qw &\qw \\
&&\qw&\qw&\targ &\ctrl{-1} &\qw&\qw&\qw&\qw&\qw&\qw&\qw&\qw&\qw&\qw&\qw &\qw&\qw&\ctrl{-1}&\targ&\qw&\qw &\qw&\qw&\qw&&&&&&&&&&&\qw&\gate{Z} &\qw &\qw \\
&\lstick{s} &\qw&\targ &\ctrl{-1} &\qw&\qw&\qw&\qw&\qw&\qw&\qw&\gate{Z} &\qw&\qw&\qw&\qw&\qw&\qw&\qw&\ctrl{-1} &\targ &\qw&\gate{P_4} &\qw &\qw &&&&&&&&&&\lstick{s}&\qw &\gate{P_4}&\qw &\qw \\
&&\qw&\ctrl{-1} &\qw&\qw&\qw&\qw &\qw&\qw&\qw&\qw&\qw&\qw&\qw&\qw&\qw&\qw&\qw&\qw&\qw&\ctrl{-1}&\qw &\qw&\qw &\qw &&&&&&&&&&&\qw&\qw&\qw&\qw 
} }\end{equation}
Consequently, by the exact same logic as that in subsection~\ref{sec:2.1}, we can ``select'' between Pauli operators corresponding to interaction terms using a circuit composed of the $\textsc{Inject}(Z)$, $\textsc{Inject-Select}(Q)$, and $\textsc{Inject-Select}(P)$ subroutines defined in subsection~\ref{sec:2.1}, along with a \textsc{Ladder} and $\textsc{Ladder}^{\dagger}$. This circuit would essentially be an extended version of that in Eq.~\eqref{selHcircuit0}, with two minor modifications. First, instead of absorbing the sign of the Pauli operator into $P_1$, as we did in subsection~\ref{sec:2.1}, we use one qubit $\ket{\mathrm{sgn}}$ in the selection register to encode the sign (with $\ket{\mathrm{sgn}} = \ket{0}$ corresponding to $+1$ and $\ket{\mathrm{sgn}} = \ket{1}$ to $-1$). We then remove the second wire in the circuit for $\textsc{Select}(Q)$ in Eq.~\eqref{SelQP}, and modify $\textsc{Inject-Select}(Q)$ accordingly. Second, to account for the possibility that different terms in the Hamiltonian may be products of different numbers of interaction operators (e.g., $a_p^\dagger a_q + \mathrm{h. c.}$ and $a_p^\dagger a_q a_r^\dagger a_s + \mathrm{h. c.}$ may both be present in a $k = 4$ Hamiltonian), we use $k$ of the qubits in the selection register as control qubits. We denote the states of these control qubits by $\ket{\textsc{i}_p}$, $\ket{\textsc{i}_q}$, $\ket{\textsc{i}_r}$, $\ket{\textsc{i}_s}$, etc. Note that the Pauli string $(P_3)_r \vec{Z}_{r,s}(P_4)_s$ on the right-hand side of Eq.~\eqref{identity4} would not be implemented if the bottom two $Z$ operators, $Q_{P_3}$, and $P_4$ are not applied on the left-hand side. It follows that by controlling the corresponding $\textsc{Inject}(Z)$, $\textsc{Inject-Select}(P)$, and $\textsc{Inject-Select}(Q)$ operators on $\ket{\textsc{i}_r}$ and $\ket{\textsc{i}_s}$, either $(P_1)_p\vec{Z}_{p,q}(P_2)_q (P_3)_r\vec{Z}_{r,s}(P_4)_q$ or $(P_1)_p\vec{Z}_{p,q}(P_2)_q$ is applied depending on the state $\ket{\textsc{i}_r}\ket{\textsc{i}_s}$. The operators associated with $p$ and $q$ are controlled as well in order to allow for the implementation of number operators, which do not transform into Pauli strings of the form in Eq.~\eqref{identity4} [cf.~Eq.~\eqref{number_operators} below]. For the example of $k = 4$, the (sub)circuit for interaction operators is the left part (labelled ``interaction operators'') of the circuit in Eq.~\eqref{hi}.

Number operators are very straightforward to implement using $\textsc{Inject}(Z)$ gates, since 
\begin{equation} \label{number_operators} n_p \mapsto \frac{1}{2}(I - Z_p)\end{equation}
under the Jordan-Wigner transformation. Therefore, to incorporate the Hamiltonian terms that involve number operators, the state of the selection register simply needs to indicate whether $Z$ operators should be applied on certain qubits (recalling that the overall sign is encoded in $\ket{\mathrm{sgn}}$). It suffices to use another $k$ of the selection register qubits as control qubits, labelling their states by $\ket{\textsc{n}_p}$, $\ket{\textsc{n}_q}$, $\ket{\textsc{n}_r}$, $\ket{\textsc{n}_s}$, etc., and control an $\textsc{Inject}(Z)$ operator on $\ket{p}$ and $\ket{\textsc{n}_p}$, another $\textsc{Inject}(Z)$ operator on $\ket{q}$ and $\ket{\textsc{n}_q}$, and so on. The full circuit for $\textsc{Select}(H)$ for any $k = 4$ Hamiltonian is shown in Eq.~\eqref{hi} below. As always, each of the $\textsc{Inject}(Z)$, $\textsc{Inject-Select}(Q)$, and $\textsc{Inject-Select}(P)$ gates can be replaced by their more efficient variants constructed in subsection~\ref{sec:2.2}.

All possible terms can be encoded by appropriately choosing the correspondence between the Pauli operators in the Jordan-Wigner encoding and the computational basis states of each subregister (and constructing $\textsc{Prepare}(\alpha)$ in a way that is consistent with this correspondence). As an explicit example, suppose that the Hamiltonian includes terms of the form $a^\dagger_p a_q n_r + \mathrm{h. c.}$, which transform to linear combinations of $(P_1)_p \vec{Z}_{p,q} (P_2)_q$ and $(P_1)_p \vec{Z}_{p,q}(P_2)_q Z_r$, with $P_1, P_2 \in \{X,Y\}$. It can be seen from Eq.~\eqref{hi} that the first type of Pauli operators are applied when $\ket{\textsc{n}_r} = \ket{0}$ and the second type are applied when $\ket{\textsc{n}_r} = \ket{1}$, with $\ket{\textsc{i}_p}\ket{\textsc{i}_q}\ket{\textsc{i}_r}\ket{\textsc{i}_s} = \ket{1}\ket{1}\ket{0}\ket{0}$ and $\ket{\textsc{n}_p}\ket{\textsc{n}_q}\ket{\textsc{n}_s} = \ket{0}\ket{0}\ket{0}$ for both. While the circuit in Eq.~\eqref{hi} implements arbitrary terms involving up to $k = 4$ spin-orbitals, it is often the case that the Hamiltonian in question only contains a few types of terms. Some of the qubits could then be removed from the selection register and the control logic could be simplified. For the molecular electronic structure Hamiltonian, which is a linear combination of  $a^\dagger_p a_q + \mathrm{h. c.}$, $a^\dagger_p a_q a^\dagger_r a_s + \mathrm{h. c.}$, $n_p$, $n_p n_q$, and $a^\dagger_p a_q n_r + \mathrm{h. c.}$~\cite{Szabo,Helgaker2000}, we would not need the qubits storing $\ket{\textsc{n}_r}$ and $\ket{\textsc{n}_s}$ and the last two $\textsc{Inject}(Z)$ gates in Eq.~\eqref{hi}, for instance. 

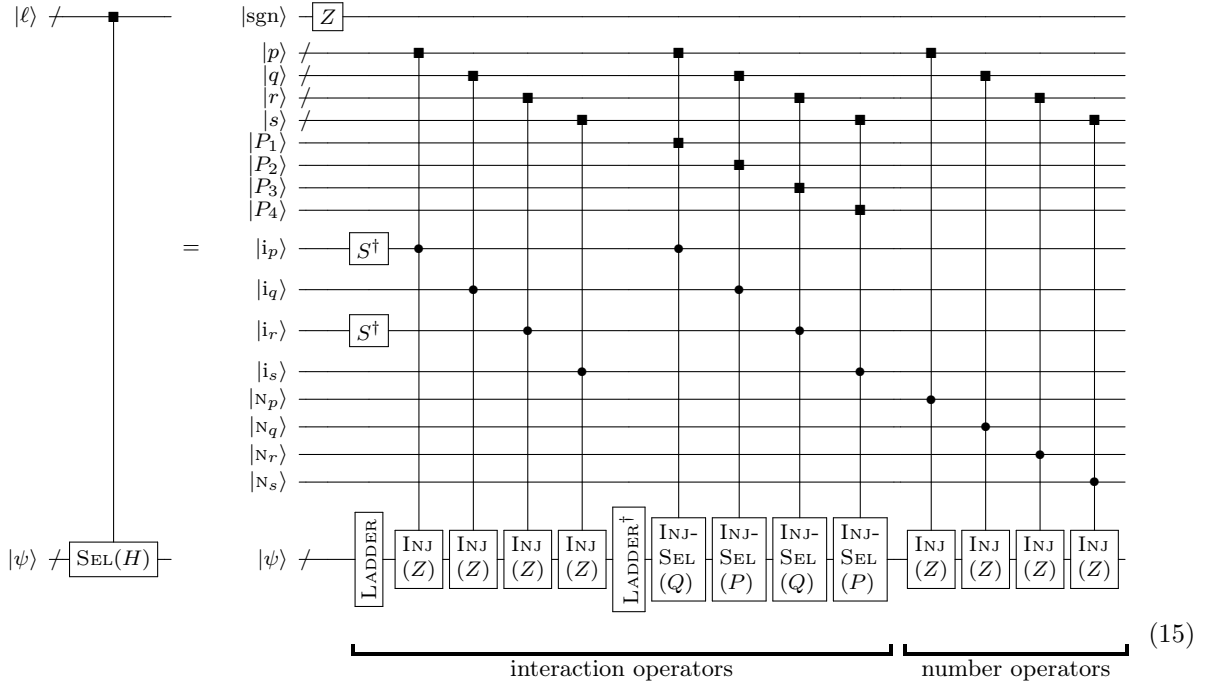
\begin{align} &{\footnotesize \Qcircuit @C=0.3em @R=1em {
&\lstick{\ket{\ell}} &/\qw &\qw &\qw\bsquare &\qw &\qw&&&&&&&&&&&&&&&&&&&\lstick{\ket{\mathrm{sgn}}} &\qw &\gate{Z} &\qw &\qw &\qw &\qw &\qw &\qw &\qw &\qw &\qw &\qw &\qw &\qw &\qw&\qw&\qw&\qw&\qw\\
&&&&&&&&&&&&&&&&&&&&&&&&&\lstick{\ket{p}} &/\qw &\qw &\qw &\qw\bsquare &\qw &\qw &\qw &\qw &\qw\bsquare &\qw &\qw &\qw &\qw &\qw &\qw\bsquare &\qw&\qw&\qw&\qw\\
&&&&&&&&&&&&&&&&&&&&&&&&&\lstick{\ket{q}} &/\qw &\qw &\qw &\qw &\qw\bsquare &\qw &\qw &\qw &\qw &\qw\bsquare &\qw &\qw &\qw &\qw&\qw&\qw\bsquare&\qw&\qw&\qw\\
&&&&&&&&&&&&&&&&&&&&&&&&&\lstick{\ket{r}} &/\qw &\qw &\qw &\qw &\qw &\qw\bsquare &\qw &\qw &\qw &\qw &\qw\bsquare &\qw &\qw&\qw &\qw&\qw&\qw\bsquare&\qw&\qw\\
&&&&&&&&&&&&&&&&&&&&&&&&&\lstick{\ket{s}} &/\qw &\qw &\qw &\qw &\qw &\qw &\qw\bsquare &\qw &\qw &\qw &\qw &\qw\bsquare &\qw &\qw&\qw&\qw&\qw&\qw\bsquare&\qw\\
&&&&&&&&&&&&&&&&&&&&&&&&&\lstick{\ket{P_1}} &\qw &\qw &\qw &\qw &\qw &\qw &\qw &\qw &\qw\bsquare &\qw &\qw &\qw &\qw&\qw &\qw&\qw&\qw&\qw&\qw\\
&&&&&&&&&&&&&&&&&&&&&&&&&\lstick{\ket{P_2}} &\qw &\qw &\qw &\qw &\qw &\qw &\qw &\qw &\qw &\qw\bsquare &\qw &\qw &\qw &\qw&\qw&\qw&\qw&\qw&\qw\\
&&&&&&&&&&&&&&&&&&&&&&&&&\lstick{\ket{P_3}} &\qw &\qw &\qw &\qw &\qw &\qw &\qw &\qw &\qw &\qw &\qw\bsquare &\qw &\qw&\qw &\qw&\qw&\qw&\qw&\qw\\
&&&&&&&&&&&&&&&&&&&&&&&&&\lstick{\ket{P_4}} &\qw &\qw &\qw &\qw &\qw &\qw &\qw &\qw &\qw &\qw &\qw &\qw\bsquare &\qw&\qw &\qw&\qw&\qw&\qw&\qw\\
&&&&&&\rstick{=}&&&&&&&&&&&&&&&&&&&\lstick{\ket{\mathrm{i}_p}} &\qw &\qw &\gate{S^\dagger} &\control\qw &\qw &\qw &\qw &\qw &\control\qw &\qw &\qw &\qw &\qw&\qw &\qw&\qw&\qw&\qw&\qw\\
&&&&&&&&&&&&&&&&&&&&&&&&&\lstick{\ket{\mathrm{i}_q}} &\qw &\qw &\qw &\qw &\control\qw &\qw &\qw &\qw &\qw &\control\qw&\qw &\qw &\qw&\qw &\qw&\qw&\qw&\qw&\qw\\
&&&&&&&&&&&&&&&&&&&&&&&&&\lstick{\ket{\mathrm{i}_r}} &\qw &\qw &\gate{S^\dagger} &\qw &\qw &\control\qw &\qw &\qw &\qw &\qw &\control\qw &\qw &\qw&\qw &\qw&\qw&\qw&\qw&\qw\\
&&&&&&&&&&&&&&&&&&&&&&&&&\lstick{\ket{\mathrm{i}_s}} &\qw &\qw &\qw &\qw &\qw &\qw &\control\qw &\qw &\qw &\qw &\qw &\control\qw &\qw&\qw &\qw&\qw&\qw&\qw&\qw\\
&&&&&&&&&&&&&&&&&&&&&&&&&\lstick{\ket{\textsc{n}_p}} &\qw&\qw&\qw&\qw&\qw&\qw&\qw&\qw&\qw&\qw&\qw&\qw&\qw&\qw &\control\qw&\qw&\qw&\qw &\qw\\
&&&&&&&&&&&&&&&&&&&&&&&&&\lstick{\ket{\textsc{n}_q}} &\qw&\qw&\qw&\qw&\qw&\qw&\qw&\qw&\qw&\qw&\qw&\qw&\qw&\qw &\qw&\control\qw&\qw&\qw&\qw\\
&&&&&&&&&&&&&&&&&&&&&&&&&\lstick{\ket{\textsc{n}_r}} &\qw&\qw&\qw&\qw&\qw&\qw&\qw&\qw&\qw&\qw&\qw&\qw&\qw&\qw&\qw&\qw&\control\qw&\qw&\qw\\
&&&&&&&&&&&&&&&&&&&&&&&&&\lstick{\ket{\textsc{n}_s}} &\qw&\qw&\qw&\qw&\qw&\qw&\qw&\qw&\qw&\qw&\qw&\qw&\qw&\qw&\qw&\qw&\qw&\control\qw &\qw\\
&\lstick{\ket{\psi}} &/\qw &\qw &\gate{\textsc{Sel}(H)}\qwx[-17] &\qw &\qw &&&&&&&&&&&&&&&&&&&\lstick{\ket{\psi}} &/\qw &\qw &\gate{\rotatebox{90}{\textsc{Ladder}}} &\gate{\injZ} \qwx[-16] &\gate{\injZ} \qwx[-15] &\gate{\injZ} \qwx[-14] &\gate{\injZ} \qwx[-13] &\gate{\rotatebox{90}{$\textsc{Ladder}^\dagger$}} &\gate{\injselQ} \qwx[-16] &\gate{\injselP} \qwx[-15] &\gate{\injselQ} \qwx[-14] &\gate{\injselP} \qwx[-13] &\qw &\qw &\gate{\injZ} \qwx[-16] &\gate{\injZ} \qwx[-15] &\gate{\injZ} \qwx[-14] &\gate{\injZ} \qwx[-13]  &\qw
}} \nonumber \\
&\hspace{11.5em} \underbracket{\hspace{20.5em}}_{\textstyle \text{\small interaction operators}} \hspace{0.25em} \underbracket{\hspace{8.5em}}_{\textstyle \text{\small number operators}} \label{hi} 
\end{align}

Thus, for arbitrary $k$, the circuit for $\textsc{Select}(H)$ can be constructed from at most $2k$ controlled-$\textsc{Inject}(Z)$, $\floor{k/2}$ controlled-$\textsc{Inject-Select}(Q)$, and $\floor{k/2}$ controlled-$\textsc{Inject-Select}(P)$, and 2 \textsc{Ladder} operators. $\textsc{Inject}(Z)$, $\textsc{Inject-Select}(Q)$, and $\textsc{Inject-Select}(P)$ are implemented using $\mathcal{O}(n)$ Clifford and $T$ gates, $\mathcal{O}(\log^2n)$ Clifford-depth, and $\mathcal{O}(\log n)$ $T$-depth, while $\textsc{Ladder}$ can be implemented using $\mathcal{O}(n)$ Clifford gates and $\mathcal{O}(\log n)$ Clifford-depth [cf.~Appendix~\ref{appendix3}]. The circuit therefore has Clifford- and $T$-count $\mathcal{O}(k n)$, Clifford-depth $\mathcal{O}(k\log^2n)$, and $T$-depth $\mathcal{O}(k\log n)$ in all cases. It is clear from Eq.~\eqref{hi} that the
selection register comprises $k\ceil{\log n} + \mathcal{O}(1)$ qubits, and that no ancillae are required. For practical applications, $k$ is usually a constant independent of $n$, in which case the total gate count is $\mathcal{O}(n)$ and the Clifford-depth and $T$-depth are $\mathcal{O}(\log^2 n)$ and $\mathcal{O}(\log n)$, respectively. The exact constant factors\footnote{As a comparison to previous work, the implementations of controlled-$\textsc{Select}(H)$ in Ref.~\cite{Babbush2018} for the molecular electronic structure Hamiltonian and for the planar Fermi-Hubbard Hamiltonian have $T$-counts of $12n + \mathcal{O}(\log n)$ and $10n + \mathcal{O}(\log n)$, respectively, and $T$-depth $\mathcal{O}(n)$. In our framework, controlled-$\textsc{Select}(H)$ can be implemented for both Hamiltonians using the same circuit, and this circuit would have $T$-count $64n + \mathcal{O}(1)$ and $T$-depth $64\ceil{\log n} + \mathcal{O}(1)$. Hence, the exponential improvement in overall circuit depth comes at the cost of only a modest constant-factor increase in the $T$-count.} in the $T$ complexity can be directly calculated using Table~\ref{table:costs}.

\section*{Acknowledgments}

The author is grateful to Isaac Kim for mentorship and helpful discussions, and to Ryan Babbush, Earl Campbell, and Craig Gidney for insightful comments on an earlier draft.

\appendix

\section{Gadgets}

\subsection{{\normalfont \textsc{SwapUp}}}  \label{appendix1}

Ref.~\cite{low2018trading} provides an ancilla-free, logarithmic-depth circuit for the $\textsc{SwapUp}$ operator defined by Eq.~\eqref{swapup}. As the ability to implement $\textsc{SwapUp}$ using only $\mathcal{O}(\log^2 n)$ Clifford-depth and $\mathcal{O}(\log n)$ $T$-depth is essential for proving our main result, we summarise the construction here. 

First, Ref.~\cite{low2018trading} observes that for any self-inverse unitary $V$ (on any number of qubits), the multi-target controlled-$V$ gate $\textsc{c}(V)_m \coloneqq \ket{0}\bra{0} \otimes I + \ket{1}\bra{1}\otimes V^{\otimes m}$ on $m$ registers can be implemented by a circuit of the form
\begin{equation} {\small \label{toggle} \Qcircuit @C=0.5em @R=0.5em {
&\qw &\ctrl{1} &\qw &&&&&&& &\qw &\ctrl{1} &\ctrl{2} &\qw &\ctrl{2} &\qw &\ctrl{3} &\qw &\ctrl{3} &\qw &\qw \\
&/\qw &\gate{V} \qwx[1] &\qw &&&&&&& &/\qw &\gate{V} &\qw &\qw &\qw &\qw &\qw &\qw &\qw &\qw &\qw \\
&/\qw &\gate{V} \qwx[1] &\qw &&\rstick{\raisebox{-2em}{=}} &&&&& &/\qw &\ctrl{1} &\targ \qwx[2] &\ctrl{1} &\targ \qwx[2] &\gate{V} &\qw &\gate{V} &\qw &\qw &\qw\\
&/\qw &\gate{V} \qwx[1] &\qw &&&&&&& &/\qw &\gate{V} &\qw &\gate{V} &\qw &\ctrl{-1} &\targ \qwx[2] &\ctrl{-1} &\targ \qwx[2] &\qw &\qw \\
&/\qw &\gate{V} \qwx[1] &\qw &&&&&&& &/\qw &\ctrl{1} &\targ &\ctrl{1} &\targ &\gate{V} &\qw &\gate{V} &\qw &\qw &\qw \\
&/\qw &\gate{V} &\qw &&&&&&& &/\qw &\gate{V} &\qw &\gate{V} &\qw &\ctrl{-1} &\targ &\ctrl{-1} &\targ &\qw &\qw
} }
\end{equation}
where \raisebox{1em}{\small $\Qcircuit @C=0.5em @R=0.5em{ &/\qw &\ctrl{1} &\qw \\ &/\qw &\gate{V} &\qw}$} indicates that a controlled-$V$ gate is applied with any one of the qubits in the first register as the control qubit. (If $m$ is even, the second wire and the topmost controlled-$V$ gate in Eq.~\eqref{toggle} are removed.) Thus, $\textsc{c}(V)_m$ can be implemented without ancilla qubits using at most $2m$ controlled-$V$ gates and $\mathcal{O}(m)$ \textsc{cnot}s in such a way that the controlled-$V$ gates are applied in $4$ layers. Ref.~\cite{low2018trading} further shows that the multi-target $\textsc{cnots}$ can be implemented in depth $\mathcal{O}(\log m)$. 

Next, consider the following circuit for $\textsc{SwapUp}$, drawn for $n = 8$:
\begin{equation}  \label{swapup_circuit} {\small \Qcircuit @C=0.7em @R=1em {
\\
\lstick{\ket{x}} &/\qw &\qw\bsquare \qwx[7] &\qw \\ 
\\ \\ &&&&&\rstick{=} \\ \\ \\ \\
\lstick{\bigotimes\limits_{y=0}^{7}\ket{\varphi_y}} &/\qw &\gate{\textsc{SwapUp}} &\qw
}
\qquad \enspace \qquad
\Qcircuit @C=0.9em @R=0.9em {
&\lstick{\ket{x_2}} &\qw &\ctrl{4} &\ctrl{5} &\ctrl{6} &\ctrl{7} &\qw&\qw&\qw&\qw&\qw&\qw&\qw  \\
&\lstick{\ket{x_1}} &\qw &\qw &\qw &\qw &\qw &\qw &\ctrl{3} &\ctrl{4} &\qw &\qw&\qw&\qw\\
&\lstick{\ket{x_0}} &\qw &\qw &\qw &\qw &\qw &\qw &\qw &\qw &\qw &\ctrl{2} &\qw&\qw\\ \\
&\lstick{\ket{\varphi_0}} &\qw &\qswap \qwx[4] &\qw &\qw &\qw &\qw &\qswap \qwx[2] &\qw &\qw &\qswap \qwx[1]&\qw &\qw &\rstick{\!\!\!\!\!\ket{\varphi_x}}\\
&\lstick{\ket{\varphi_1}} &\qw &\qw &\qswap\qwx[4] &\qw &\qw &\qw &\qw &\qswap \qwx[2] &\qw &\qswap &\qw &\qw\\
&\lstick{\ket{\varphi_2}} &\qw &\qw &\qw &\qswap\qwx[4] &\qw &\qw &\qswap &\qw&\qw&\qw&\qw&\qw\\
&\lstick{\ket{\varphi_3}} &\qw &\qw &\qw &\qw &\qswap\qwx[4] &\qw &\qw &\qswap &\qw&\qw&\qw&\qw\\
&\lstick{\ket{\varphi_4}} &\qw &\qswap &\qw&\qw&\qw&\qw&\qw&\qw&\qw&\qw&\qw&\qw\\
&\lstick{\ket{\varphi_5}} &\qw &\qw &\qswap &\qw&\qw&\qw&\qw&\qw&\qw&\qw&\qw&\qw\\
&\lstick{\ket{\varphi_6}} &\qw &\qw &\qw &\qswap &\qw&\qw&\qw&\qw&\qw&\qw&\qw&\qw\\
&\lstick{\ket{\varphi_7}} &\qw &\qw &\qw &\qw &\qswap &\qw&\qw&\qw&\qw&\qw&\qw&\qw
} }
\end{equation}
Here, $x_2x_1x_0$ is the binary representation of $x$, where $x_0$ is the least significant bit. The basic idea is that after the controlled-\textsc{swap} gates controlled on the qubit storing $\ket{x_2}$ are applied, the states $\ket{\varphi_{y}}$ with $y_2 = x_2$ are on the first four qubits of the second register (writing $y$ in binary as $y_2 y_1 y_0$). After the controlled-\textsc{swap}s  controlled on $\ket{x_1}$ are applied, the states $\ket{\varphi_y}$ with $y_2 = x_2$ and $y_1 = x_1$ are on the first two qubits, and finally the controlled-\textsc{swap} controlled on $\ket{x_0}$ ensures that $\ket{\varphi_x}$ ends up on the very first qubit. In general, if the target register has $n$ qubits (where $n$ is not necessarily a power of $2$) and $x_{\ceil{\log n} -1}\dots x_1x_0$ is the binary representation of $x$, $n - 2^{\ceil{\log n} - 1}$ controlled-\textsc{swap}s are controlled on $\ket{x_{\ceil{\log n - 1}}}$, followed by $2^j$ controlled-\textsc{swap}s controlled on $\ket{x_j}$ for each $j \in \{0, \dots, \ceil{\log n} - 2\}$ in descending order. 

Note that for each $j \in \{0, \dots, \ceil{\log n} - 1\}$, the controlled-\textsc{swap} gates controlled on $\ket{x_j}$ target disjoint pairs of qubits. Since $\textsc{swap}$ is self-inverse, we can apply Eq.~\eqref{toggle} with $V = \textsc{swap}$ and $m \leq 2^j$ (specifically, $m = 2^j$ for $j \in \{0,\dots, \ceil{\log n} - 2\}$ and $m = n - 2^{\ceil{\log n} - 1}$ for $j = \ceil{\log n} - 1$) to replace all of the controlled-\textsc{swap}s controlled on $\ket{x_j}$ in Eq.~\eqref{swapup_circuit} with at most $2\cdot 2^j$ controlled-\textsc{swap}s applied in $4$ layers, along with $\mathcal{O}(2^j)$ \textsc{cnot}s applied in $\mathcal{O}(j)$ layers. Summing over $j$, the total number of controlled-\textsc{swaps} is 
\[ 2\left(n - 2^{\ceil{\log n} - 1}\right) + \sum_{j=0}^{\ceil{\log n} - 2} 2\cdot 2^j= 2(n-1) \]
and these are applied in $4\ceil{\log n}$ layers. 
Each controlled-\textsc{swap} can be decomposed into one \textsc{Toffoli} and two \textsc{cnot} gates, and each \textsc{Toffoli} has $T$-count $7$ and $T$-depth $4$. Therefore, $\textsc{SwapUp}$ has $T$-count $14(n-1)$ and $T$-depth $16\ceil{\log n}$. 
The total number of Clifford gates (including the multi-target \textsc{cnot}s in Eq.~\eqref{toggle} and the Cliffords in the decomposition of each controlled-\textsc{swap}) is $\mathcal{O}(n)$, and the total Clifford-depth is $\mathcal{O}(\log^2n)$. 

\subsection{{\normalfont $\textsc{SwapUp}^*$}} \label{appendix2}

In subsection~\ref{sec:2.2}, we use the fact that a certain phase-incorrect version $\textsc{SwapUp}^*$ of \textsc{SwapUp} is less costly than \textsc{SwapUp} to reduce the $T$-count and $T$-depth by constant multiplicative factors. $\textsc{SwapUp}^*$ is obtained by replacing each controlled-\textsc{swap} gate in Eq.~\eqref{swapup_circuit} with a phase-incorrect version of controlled-\textsc{swap}: 
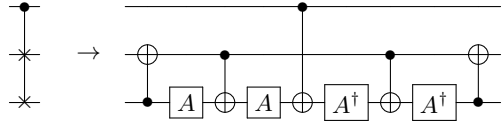
\[ {\small \Qcircuit @C=0.5em @R=0.5em @!R {
&\ctrl{1} &\qw & &&&&&&&\qw &\qw &\qw  &\qw &\ctrl{2} &\qw &\qw &\qw &\qw &\qw \\
&\qswap \qwx[1] &\qw &&\rstick{\to} &&&&& &\targ &\qw &\ctrl{1} &\qw &\qw &\qw &\ctrl{1} &\qw &\targ &\qw \\
&\qswap &\qw &&&&&&&&\ctrl{-1} &\gate{A} &\targ &\gate{A} &\targ &\gate{A^\dagger} &\targ&\gate{A^\dagger}  &\ctrl{-1} &\qw
}}
\]
where $A \coloneqq e^{i(\pi/8)Y} = S^\dagger HTHS$ (here, $H$ denotes the Hadamard gate). The circuit on the right-hand side implements an operator whose action on computational basis states differs from controlled-\textsc{swap} only in that $\ket{100}$ is mapped to $-\ket{100}$. Hence, since controlled-\textsc{swap} merely permutes the computational basis states, $\textsc{SwapUp}^*$ acts the same as \textsc{SwapUp} on computational basis states up to sign. As observed in Ref.~\cite{low2018trading}, any sequence of these phase-incorrect controlled-\textsc{swap} operators that target disjoint pairs of qubits can be parallelised such that the $A$ and $A^\dagger$ gates are applied in four layers, and the \textsc{cnot}s in $\mathcal{O}(\log n)$ layers. It follows that $\textsc{SwapUp}^{*}$ has $T$-count $4(n-1)$ and $T$-depth $4\ceil{\log n}$ (and Clifford-count $\mathcal{O}(n)$ and Clifford-depth $\mathcal{O}(\log^2 n)$). 

\subsection{\normalfont \textsc{Ladder}} \label{appendix3}

The \textsc{Ladder} operator, defined in Eq.~\eqref{ladder} by a cascade of $n-1$ \textsc{cnot} gates, can equivalently be implemented using a logarithmic-depth circuit, as pointed out by Gidney~\cite{gidney}. By Eq.~\eqref{ladder}, \textsc{Ladder} acts on computational basis states as 
\[ \textsc{Ladder}: \ket{z_0}\ket{z_1} \dots \ket{z_{n-1}} \mapsto \ket{z_0 \oplus \dots \oplus z_{n-1}}\ket{z_1 \oplus \dots \oplus z_{n-1}} \dots \ket{z_{n-1}}. \] It is easily verified that the same transformation can be realised by arranging \textsc{cnot} gates in a tree-like structure, shown below for $n = 8$:
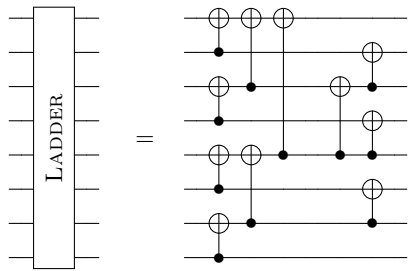
\[ \small { \Qcircuit @C=0.5em @R=0.5em {
&\qw &\multigate{7}{\rotatebox{90}{\textsc{Ladder}}} &\qw &\qw &&&&&&&&\qw &\targ &\targ &\targ &\qw&\qw&\qw &\qw &\qw &\qw \\
&\qw &\ghost{\rotatebox{90}{\textsc{Ladder}}} &\qw &\qw &&&&&&&&\qw &\ctrl{-1} &\qw &\qw &\qw &\qw &\qw&\targ &\qw &\qw \\
&\qw &\ghost{\rotatebox{90}{\textsc{Ladder}}} &\qw &\qw &&&&&&&&\qw &\targ &\ctrl{-2} &\qw &\qw &\qw&\targ &\ctrl{-1} &\qw &\qw \\
&\qw &\ghost{\rotatebox{90}{\textsc{Ladder}}} &\qw &\qw &&\rstick{\raisebox{-2em}{$=$}} &&&&&&\qw &\ctrl{-1} &\qw &\qw&\qw&\qw&\qw &\targ &\qw &\qw\\
&\qw &\ghost{\rotatebox{90}{\textsc{Ladder}}} &\qw &\qw &&&&&&&&\qw &\targ &\targ &\ctrl{-4}  &\qw&\qw &\ctrl{-2} &\ctrl{-1} &\qw &\qw\\
&\qw &\ghost{\rotatebox{90}{\textsc{Ladder}}} &\qw &\qw &&&&&&&&\qw &\ctrl{-1} &\qw &\qw &\qw  &\qw&\qw &\targ &\qw &\qw\\
&\qw &\ghost{\rotatebox{90}{\textsc{Ladder}}} &\qw &\qw &&&&&&&&\qw &\targ &\ctrl{-2} &\qw &\qw &\qw&\qw &\ctrl{-1} &\qw &\qw\\
&\qw &\ghost{\rotatebox{90}{\textsc{Ladder}}} &\qw &\qw &&&&&&&&\qw &\ctrl{-1} &\qw &\qw &\qw &\qw &\qw&\qw &\qw &\qw\\
}
}
\]
(For $n$ that is not a power of $2$, the circuit for $\textsc{Ladder}$ can be obtained by starting with the circuit for the next largest power of $2$, then removing all of the \textsc{cnot}s supported on qubits that are out of range.) Thus, a circuit $\textsc{Ladder}$ can be constructed using $\mathcal{O}(n)$ \textsc{cnot}s applied in $\mathcal{O}(\log n)$ layers.

\bibliographystyle{unsrtnat}
\bibliography{bib.bib} 

\end{document}